# Chaotic Encryption for 10-Gb Ethernet Optical Links

Adrián Pérez-Resa, Miguel Garcia-Bosque, Carlos Sánchez-Azqueta, and Santiago Celma

*Abstract*—In this paper, a new physical layer encryption method for optical 10-Gb Ethernet links is proposed. Necessary modifications to introduce encryption in Ethernet 10GBase-R standard have been considered. This security enhancement has consisted of a symmetric streaming encryption of the 64b/66b data flow at physical coding sublayer level thanks to two keystream generators based on a chaotic algorithm. The overall system has been implemented and tested in a field programmable gate array. Ethernet traffic has been encrypted, transmitted, and decrypted over a multimode optical link. Experimental results are analyzed concluding that it is possible to cipher traffic at this level and hide the complete Ethernet traffic pattern from any passive eavesdropper. In addition, no overhead is introduced during encryption, getting no losses in the total throughput.

*Index Terms*—Ethernet, 10GBASE-R, cryptography, stream cipher, skew tent map.

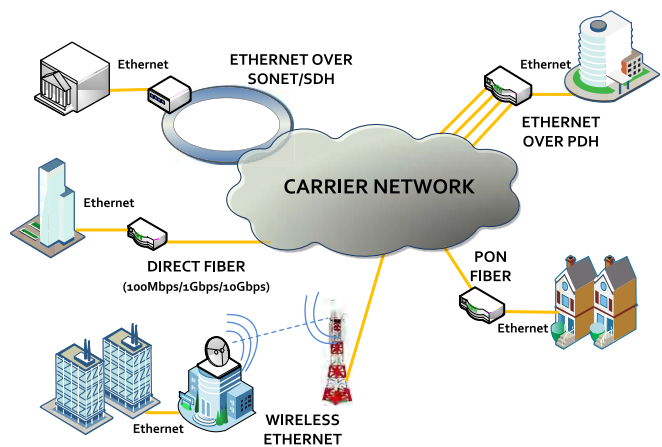

Fig. 1. Simple example of different access technologies for a CEN (Carrier Ethernet Network).

## I. INTRODUCTION

NOWADAYS, thanks to advances in optical communication technologies it is possible to meet the ever-increasing data traffic demand of modern networks. It is estimated that IP traffic in data centers will reach 15.3 Zettabytes per year by 2020 [1].

On the other hand, security and confidentiality in communications networks are important fields of study. In the specific case of optical networks, some studies were carried out in the past to discern whether the degradation of a link or failure of a service is due to a natural network defect or an external attack [2], [3]. In fact, threat analysis in the physical layer is considered crucial for secure communications [4], [5]. However, nowadays there are simple methods for intercepting the signal without appreciably interfering in optical communication [6]. Such methods use simple fiber coupling devices and electro-optical converters and numerous examples are easily available for the public on the Internet.

In a layered communication model, such as OSI (Open System Interconnection) or TCP/IP (Transmission Control Protocol/Internet Protocol), 'in-flight' encryption can be carried out at different communication layers. For example, well known encryption methods for layer 2 and layer 3 are the protocols MACsec [7] and IPsec [8] respectively.

At physical layer, there are security solutions directly related to the optical technology such as QKD (Quantum Key Distribution) [9] or OCDM (Optical Code-Division Multiplexing) [10], or related to payload encryption in some optical protocols as OTN (Optical Transport Network) standard [11]. Encryption at physical layer brings some benefits such as minimum latency and zero overhead, achieving maximum link efficiency. In fact, commercial OTN equipment performing encryption at line rate is able to achieve a 100% throughput in contrast with encryption at other communication layers [12]. As an example, in IPsec the inherent overhead introduced during encryption reduces the overall throughput between 20% and 90% of the maximum achievable [13].

Today one of the most used technologies for the access to optical transport networks is Ethernet. As shown in Fig. 1, some access technologies in CENs (Carrier Ethernet Networks) 'last mile' are Ethernet over Fiber (Direct Fiber, SONET/SDH, PON), Ethernet over PDH, Wireless Ethernet, etc. [14].

10GBase-R is one of the most widely used physical layer standards for Ethernet at 10 Gbps in optical links, but nowadays there is no mechanism or standard providing physical layer security on it. For this reason, the main motivation of this work is to propose and implement the necessary modifications

Manuscript received February 13, 2018; revised July 10, 2018 and August 24, 2018; accepted August 25, 2018. This work was supported in part by MINECO-FEDER under Grant TEC2014-52840-R and Grant TEC2017-85867-R. The work of M. Garcia-Bosque was supported by FPU Fellowship under Grant FPU14/03523. This paper was recommended by Associate Editor Guido Masera. *(Corresponding author: Adrián Pérez-Resa.)*

The authors are with the Electronic and Communications Engineering Department, University of Zaragoza, 50009 Zaragoza, Spain (e-mail: aprz@unizar.es; mgbosque@unizar.es; csanaz@unizar.es; scelma@unizar.es).

Color versions of one or more of the figures in this paper are available online at http://ieeexplore.ieee.org.

Digital Object Identifier 10.1109/TCSI.2018.2867918

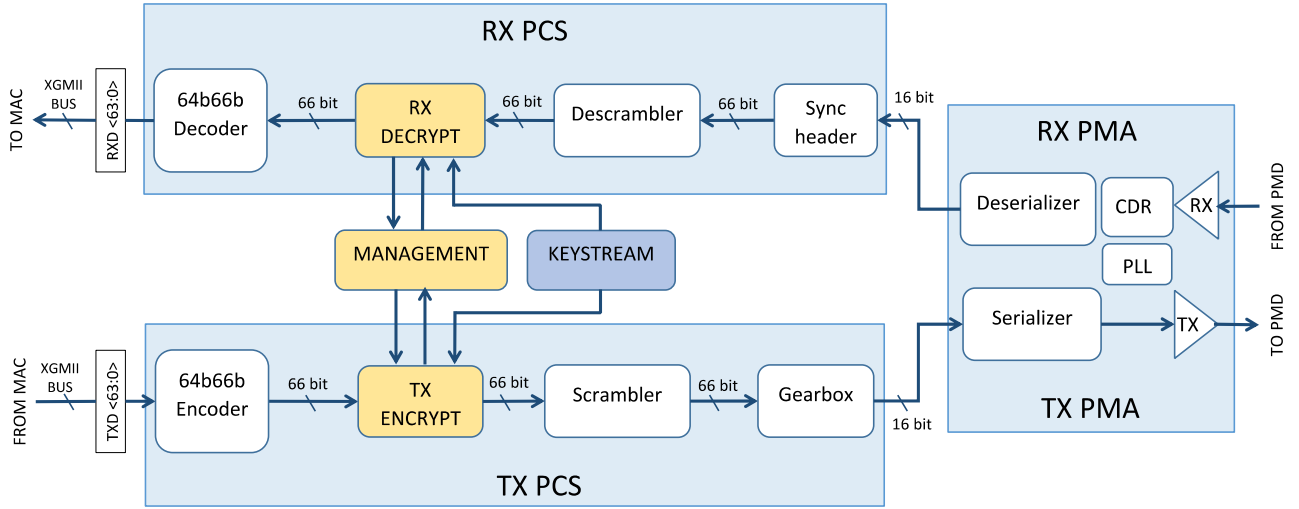

Fig. 2. PCS structure and the proposed encryption function 10G-PHYsec.

to introduce encryption in its PCS (Physical Coding Sublayer) level. We have called this method 10G-PHYsec.

An encryption system at physical layer of Ethernet provides the mentioned advantages such as no overhead, low latency and also obfuscation of customer data traffic patterns, which results in an overall improvement of security. An example of the same property is proved at a lower transmission speed in an Ethernet interface using 1000Base-X physical layer [15]. In this case the encryption mechanism is completely different, as 8b/10b encoding is used, consisting of an additive stream cipher where modulo-267 addition is performed between the keystream and plaintext.

The article is organized as follows: Section II explains the general structure of the encryption system in the Ethernet physical layer 10GBase-R, Section III explains the encryption infrastructure composed by the stream cipher operation and the synchronization mechanism between TX (transmitter) and RX (receiver), Section IV deals with keystream generation of the proposed stream cipher; Section V details the system implementation and test results and finally in Section VI conclusions obtained in this work are given.

## II. PCS Layer Encryption

In Ethernet standards, the physical layer is usually divided into three sublayers with different functionalities: PCS (Physical Coding Sublayer), PMD (Physical Medium Dependent) and PMA (Physical Medium Attachment). PCS sublayer performs functions such as link establishment, data encoding, scrambling, synchronization, clock rate adaptation, etc.

As many high speed standards in optical Ethernet, a baseband serial data transmission is carried out while clock frequency information is embedded in the serial bitstream itself. At PMA sublayer a CDR (Clock and Data Recovery) circuit is responsible for extracting timing information from the data stream. As a critical part of a typical transceiver for high speed communications many researches have been carried out about this kind of circuits [16], [17]. In order to facilitate the work of the CDR circuit, information is usually encoded in such a way that a good transition density and short run length are achieved. Also a DC-balanced serial data stream must be guaranteed by getting a good disparity, which is important for some transmission media, as optical links.

In the case of 10GBase-R standard, 64b/66b encoding is used at PCS sublayer. This type of block-line encoding achieves the properties set out before in a statistical way thanks to the scrambling of the bitstream. The basic structure of the PCS sublayer in 10GBase-R standard is shown in Fig. 2, where its main function blocks such as 64b/66b encoder/decoder and scrambler/descrambler are shown.

In order to encrypt physical layer when block-line encoding is used, it is necessary to preserve the mentioned encoder properties, therefore the location of the cipher in the datapath must be taken into account. In this work we propose to carry out encryption at the input of the scrambler and decryption at the output of the descrambler as shown in Fig. 2

Unlike other encryption mechanisms such as the MACsec, where the encrypted data are directly the Ethernet packets, in this work, the 64b/66b data flow is directly encrypted thanks to a stream cipher based on a chaotic algorithm. The main advantages of this method is that there is no throughput loss and that not only data packet content is encrypted but also the link activity or Ethernet data traffic pattern. By encrypting at 64b/66b block level, both data and control blocks are obfuscated, such as start and end frame blocks or control blocks with IDLE control characters inside, making frame pattern indecipherable.

## III. Encryption Infrastructure

The encryption infrastructure is shown in Fig. 3. This structure is composed of three blocks, the TX_ENCRYPT, RX_DECRYPT and the MANAGEMENT module.

### A. Encryption Function

The stream cipher operation is performed by the CIPHER_OP_TX and CIPHER_OP_RX modules in Fig. 3.

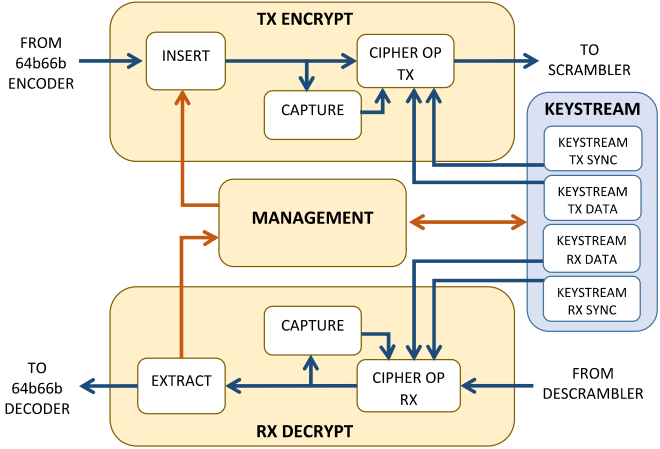

Fig. 3. Encryption infrastructure for 10G-PHYsec function.

The encoded blocks after the 64b/66b encoder are composed of a two-bit synchronization header plus a 64-bit payload, forming a 66-bit block. On the one hand, the 66-bit block type depends on the type field inside the block payload and the value of the 2-bit block synchronization header, then both, block payload and synchronization header are necessary to be encrypted to mask the block type. On the other hand, the main purpose of the synchronization header is to guarantee 66-bit block alignment while limiting the run length to 66 bits. This header only has two possible values ('01' or '10'), which produces a transition between 0 and 1 every 66 bits. Thus, after encryption, it is necessary to preserve this bit transition.

The stream cipher operation is shown in Fig. 4. Both modules CIPHER_OP_TX and CIPHER_OP_RX perform the same operation. Firstly, the 66-bit block, D_IN, is split in two parts, the synchronization header and the block payload. The block payload is directly encrypted by performing an XOR operation between its 64 bits and a 64-bit keystream data sequence, called 'keystream data' in Fig. 4. The 2-bit synchronization header is mapped to values '0' or '1' depending, respectively, on whether it is equal to '01' or '10'. Then, the mapped value is encrypted performing an XOR operation with a 1-bit keystream sequence called 'keystream sync'. After encryption, this new value is reverse mapped resulting in a new header '01' or '10' depending on whether it is '0' or '1'. In this way, the transition between 0 and 1 every 66 bits is guaranteed. Finally, the new synchronization header is concatenated with the encrypted block payload resulting in the output D_OUT, which is sent to the scrambler when encrypting or to the 64b/66b decoder when decrypting.

### B. Encryption Synchronization

It is necessary to activate and deactivate the keystream generators and the encryption operation synchronously between transmitter and receiver. For that purpose, the MANAGEMENT module in Fig. 3 performs the insertion and extraction of two management sequences into the 64b/66b block stream. One of these sequences is used for the encryption activation and the other one for deactivation.

TABLE I
64B/66B BLOCK FORMATS IN 10GBASE-R STANDARD

| Input Data | Sync | Block Payload | | | | | | | |
|---|---|---|---|---|---|---|---|---|---|
| Bit Position: | 0 1 | 2 | | | | | | | 65 |
| **Data Block Format:** | | | | | | | | | |
| $D_0 D_1 D_2 D_3/D_4 D_5 D_6 D_7$ | 01 | $D_0$ | $D_1$ | $D_2$ | $D_3$ | $D_4$ | $D_5$ | $D_6$ | $D_7$ |
| **Control Block Formats:** | | Block Type | | | | | | | |
| $C_0 C_1 C_2 C_3/C_4 C_5 C_6 C_7$ | 10 | 0x1e | $C_0$ | $C_1$ | $C_2$ | $C_3$ | $C_4$ | $C_5$ | $C_6$ | $C_7$ |
| $C_0 C_1 C_2 C_3/O_4 D_5 D_6 D_7$ | 10 | 0x2d | $C_0$ | $C_1$ | $C_2$ | $C_3$ | $O_4$ | $D_5$ | $D_6$ | $D_7$ |
| $C_0 C_1 C_2 C_3/S_4 D_5 D_6 D_7$ | 10 | 0x33 | $C_0$ | $C_1$ | $C_2$ | $C_3$ | | $D_5$ | $D_6$ | $D_7$ |
| $O_0 D_1 D_2 D_3/S_4 D_5 D_6 D_7$ | 10 | 0x66 | $D_1$ | $D_2$ | $D_3$ | $O_0$ | | $D_5$ | $D_6$ | $D_7$ |
| $O_0 D_1 D_2 D_3/O_4 D_5 D_6 D_7$ | 10 | 0x55 | $D_1$ | $D_2$ | $D_3$ | $O_0$ | $O_4$ | $D_5$ | $D_6$ | $D_7$ |
| $S_0 D_1 D_2 D_3/D_4 D_5 D_6 D_7$ | 10 | 0x78 | $D_1$ | $D_2$ | $D_3$ | $D_4$ | | $D_5$ | $D_6$ | $D_7$ |
| $O_0 D_1 D_2 D_3/C_4 C_5 C_6 C_7$ | 10 | 0x4b | $D_1$ | $D_2$ | $D_3$ | $O_0$ | $C_4$ | $C_5$ | $C_6$ | $C_7$ |
| $T_0 C_1 C_2 C_3/C_4 C_5 C_6 C_7$ | 10 | 0x87 | | $C_1$ | $C_2$ | $C_3$ | $C_4$ | $C_5$ | $C_6$ | $C_7$ |
| $D_0 T_1 C_2 C_3/C_4 C_5 C_6 C_7$ | 10 | 0x99 | $D_0$ | | $C_2$ | $C_3$ | $C_4$ | $C_5$ | $C_6$ | $C_7$ |
| $D_0 D_1 T_2 C_3/C_4 C_5 C_6 C_7$ | 10 | 0xaa | $D_0$ | $D_1$ | | $C_3$ | $C_4$ | $C_5$ | $C_6$ | $C_7$ |
| $D_0 D_1 D_2 T_3/C_4 C_5 C_6 C_7$ | 10 | 0xb4 | $D_0$ | $D_1$ | $D_2$ | | $C_4$ | $C_5$ | $C_6$ | $C_7$ |
| $D_0 D_1 D_2 D_3/T_4 C_5 C_6 C_7$ | 10 | 0xcc | $D_0$ | $D_1$ | $D_2$ | $D_3$ | | $C_5$ | $C_6$ | $C_7$ |
| $D_0 D_1 D_2 D_3/D_4 T_5 C_6 C_7$ | 10 | 0xd2 | $D_0$ | $D_1$ | $D_2$ | $D_3$ | $D_4$ | | $C_6$ | $C_7$ |
| $D_0 D_1 D_2 D_3/D_4 D_5 T_6 C_7$ | 10 | 0xe1 | $D_0$ | $D_1$ | $D_2$ | $D_3$ | $D_4$ | $D_5$ | | $C_7$ |
| $D_0 D_1 D_2 D_3/D_4 D_5 D_6 T_7$ | 10 | 0xff | $D_0$ | $D_1$ | $D_2$ | $D_3$ | $D_4$ | $D_5$ | $D_6$ | |

TABLE II
LINK FAULT SIGNALING ORDERED SETS AND
NEW PROPOSED ORDERED SETS

| LANE 0 | LANE 1 | LANE 2 | LANE 3 | DESCRIPTION |
|---|---|---|---|---|
| Sequence | 0x00 | 0x00 | 0x00 | Reserved |
| Sequence | 0x00 | 0x00 | 0x01 | Local Fault |
| Sequence | 0x00 | 0x00 | 0x02 | Remote Fault |
| Sequence | 0x00 | 0x00 | 0x03 | Link Interruption |
| **Sequence** | **0x00** | **0x00** | **0x04** | **Cipher ON** |
| **Sequence** | **0x00** | **0x00** | **0x05** | **Cipher OFF** |
| Sequence | ≥0x00 | ≥0x00 | ≥0x06 | Reserved |

In order to maintain the concordance with the standard, the format of the implemented control sequence is equivalent to that of an ordered set. Among available 64b/66b block types, the one selected for this work is 0x55. As shown in Table I, inside this block type two consecutive ordered sets can be transmitted.

On the other hand, the possible ordered sets established in the standard are between values 0x00 and 0x03 and they are used for link fault signaling. Since values above 0x03 are out of use and are reserved for future standardization, in this work the new ordered sets for encryption control signaling have been defined as shown in Table II. They have been called 'Cipher ON' and 'Cipher OFF', and their values are 0x04 and 0x05, respectively.

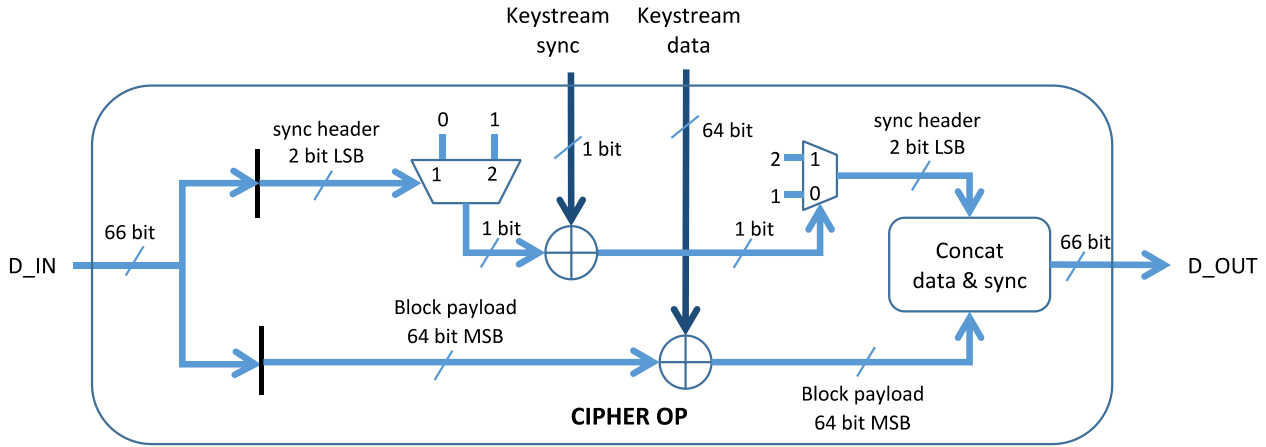

Fig. 4. Stream cipher operation.

The final encryption sequence for enabling the encryption consists of the 0x55 block type filled with two consecutive 'Cipher ON' ordered sets. In this work this block is called 'Cipher_ON block'. For disabling encryption the sequence is the same but using two 'Cipher OFF' ordered sets and the resulting block is called 'Cipher_OFF block'.

For carrying out the insertion of these new control sequences the structure shown in Fig. 3 has been used. This infrastructure allows carrying out the generation, insertion and capture of a future set of messages that extends the encryption management functionality, such as key refresh or other options between transmitter and receiver. The basic operation of this infrastructure is described below.

The CIPHER_OP_TX module is responsible for implementing the encryption operation. Initially, it is disabled, allowing the 64b/66b blocks to be transparently passed from the encoder to the scrambler. To start the encryption of the 64b/66b block stream, the MANAGEMENT module acts on the INSERT module to send the encryption start message. This message is inserted in 64b/66b block stream by replacing one 0x1E type block filled with IDLE control characters /I/ with the new Cipher_ON block. When CAPTURE module detects the presence of a Cipher_ON block it enables the CIPHER_OP_TX module and TX keystream generators (KEYSTREAM TX SYNC and KEYSTREAM TX DATA in Fig. 3), which starts the encryption process after Cipher_ON block has been transmitted.

In the receiver, the module CIPHER_OP_RX is in charge of performing the decryption operation. Like the transmitter, it is initially inactive, enabling 64b/66b blocks to be transparently passed from the descrambler to the decoder. When the CAPTURE module receives the Cipher_ON block, CIPHER_OP_RX module and RX keystream generator modules (KEYSTREAM RX SYNC and KEYSTREAM RX DATA in Fig. 3) are enabled, starting to decrypt the 64b/66b data flow after Cipher_ON block. Subsequently, the control sequence Cipher_ON is extracted from the data stream in the EXTRACT module, which replaces it with a 0x1e type block filled with /I/ control characters (reverse operation of the INSERT module). Thanks to this procedure, the keystream generators in TX and RX are synchronized and data can be deciphered correctly.

In order to disable the encryption, the same process is used, but sending Cipher_OFF blocks instead of Cipher_ON.

## IV. ENCRYPTION KEYSTREAM

The keystream generator is responsible for generating a pseudorandom sequence to carry out the encryption of the signal. In this work the generation of this pseudorandom sequence is based on a chaotic algorithm.

As the encryption operation consists of two XOR operations, one for block payload and another one for the synchronization header, two keystream generators are necessary. Both generators have different output widths, the generator for block payload encryption produces a 64-bit sequence, while the keystream generator for synchronization header is only 1-bit wide, as shown in Fig. 4.

Next, a description of the keystream generator is given, and also some security considerations are discussed, such as the achieved keystream randomness or the key space used.

### A. Chaotic Pseudorandom Bitstream Generator

In the last decades chaotic systems have been taken into account for the design of new cryptographic systems. Their high sensitivity to the initial conditions or control parameters (also called 'butterfly effect') and their unpredictable pseudo-random orbits are the most well-known characteristics of chaotic systems. These characteristics are very similar to those that a good cryptographic algorithm should have [18]. All these considerations have given rise to new private and public encryption proposals [19], [20].

A variant of the chaotic map called STM (Skew-Tent-Map) has been used in this work to generate the pseudorandom bitstream, used as keystream of the proposed cryptographic system. Based on previous work by our group, this chaotic map has already been theoretically studied in [21] and [22]. The hardware block diagram of the basic STM algorithm is shown in Fig. 5 and its equations are shown below:

$$f(x_i) = x_{i+1} = \begin{cases} x_i/\gamma, & x_i \in [0, \gamma] \\ (1 - x_i)/(1 - \gamma), & x_i \in (\gamma, 1] \end{cases} \quad (1)$$

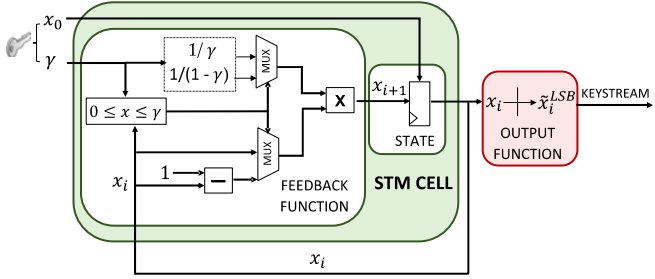

Fig. 5. Skew Tent Map block diagram.

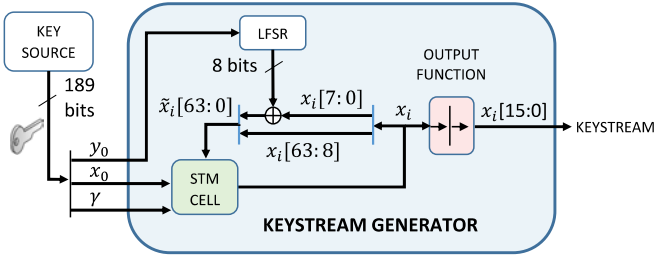

Fig. 6. Basic keystream generator. The 189-bit key is the concatenation of the three parameters ($y_0, x_0, \gamma$).

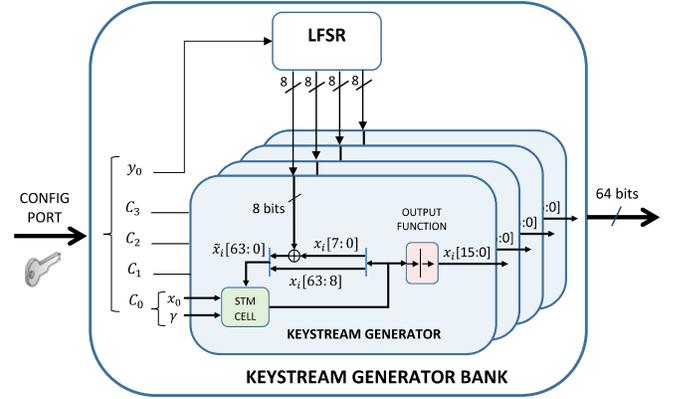

Fig. 7. 64-Bit Keystream Generator for 64b/66b block payload.

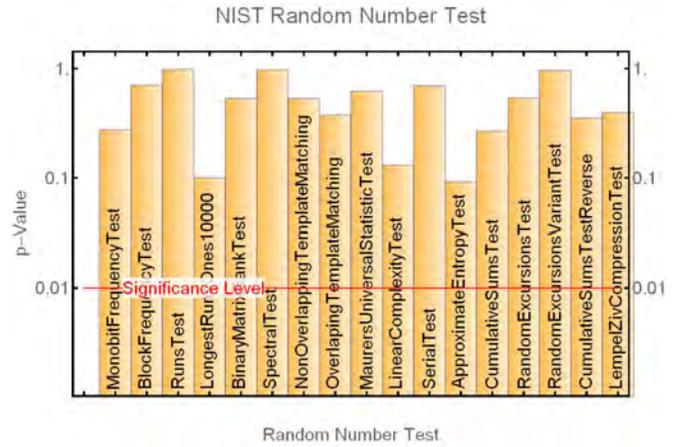

Fig. 8. NIST test results for a bitstream generated using STM-LFSR algorithm.

where its control parameter $\gamma$ and initial state $x_0$ are real numbers whose values are included in the interval (0, 1). One of the main advantages of this chaotic map is that it maintains its chaoticity for all values of $\gamma$ and $x_0$, which makes it more adequate for cryptographic applications than other maps whose parameter space contains periodic windows [18].

One important problem of digital implementations of chaotic maps under finite-precision and fixed-point arithmetic is that their dynamical properties are degraded with respect to their original continuous versions [23]. As studied in [21] and [22] an LFSR (Linear Feedback Shift Register) can be used to improve the digital implementation of the chaotic map shown in (1), reducing the mentioned dynamical degradation. Thanks to this modification it is possible to obtain an important increase in the keystream period and therefore achieve better randomness [24].

The basic keystream generator used in this work is shown in Fig. 6. A 61-step LFSR has been added to the Skew Tent Map module. The key for this basic keystream generator consists of the initial state and control parameter ($\gamma$, $x_0$), of the STM cell and the initial state of the LFSR ($y_0$).

The STM cell has been implemented with an internal 64-bit state whose output is the vector $x_i[63:0]$. The improvement related to the LFSR is to perform the XOR operation between the least significant 8 bits of LFSR and the STM cell outputs. The state returned to the STM cell will no longer be the previous state calculated by it, but a new state, $\tilde{x}_i$ to which a small noise generated by the LFSR has been added thanks to the XOR operation. The generator output function consists of taking the 16 least significant bits of $x_i$ and discarding the rest.

For the payload block encryption, a 64-bit keystream sequence is necessary, thus a bank of basic keystream generators has been built. The bank consists of four 16-bit generators whose outputs are concatenated to give a 64-bit output. This structure is shown in Fig. 7.

For the synchronization header encryption, as only a 1-bit keystream sequence is needed the basic STM cell in Fig. 6 has been used, but in this case only the least significant bit of $x_i$ is taken as output.

### B. Keystream Output Analysis

The keystream obtained at the output of both generators must be analyzed in order to test their randomness, as one important requirement for a secure stream cipher is that their keystream must be undistinguishable from a truly random sequence. In order to check this property, sequences generated by the basic generator in Fig. 6 were subjected to the NIST (National Institute of Standards and Technology) SP 800-22 battery of test [25]. Both generators passed these tests. An example is shown in Fig. 8, where a particular sequence is tested.

### C. Security Considerations: Key Space

One of the security considerations that has to be taken into account is the key space size [18]. Some recommendations establish 112 bits as minimum key size, as NIST recommendation [26], others as ENISA (European Union Agency for Network and Information Security) report [27] recommends 128 or 256 bits for mid-term and long-term security respectively.

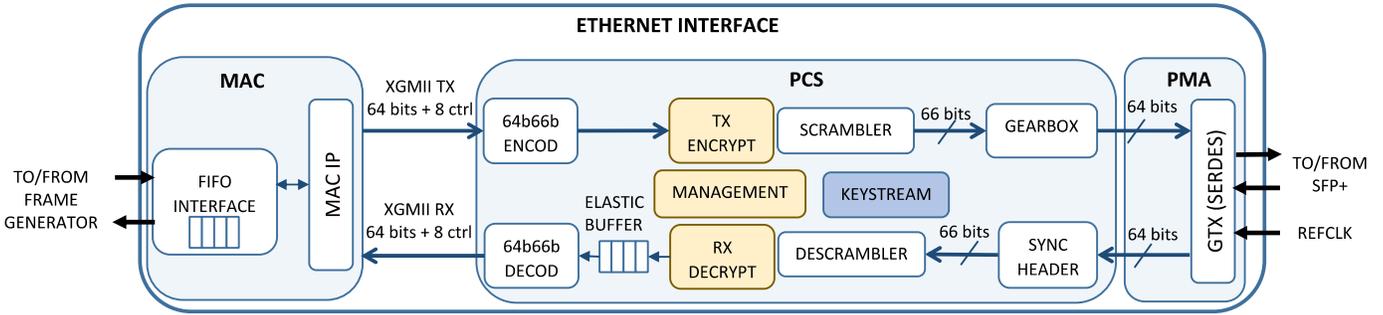

Fig. 9. Scheme of FPGA implementation for 10G Ethernet interface with encryption function.

For the basic keystream generator in Fig. 6, the complete key is composed by two 64-bit parameters ($\gamma$, $x_0$) and one 61-bit parameter $y_0$. The resulting key is 189-bit length, which can be considered good enough to guarantee security against brute-force attacks.

### D. Security Considerations: Sensitivity

Sensitivity to the initial conditions and control parameters is a well-known characteristic of chaotic maps. It leads to the desired diffusion property that any cryptosystem must have under small changes in the secret key. While in the proposed cryptosystem, the STM cell meets this property, sensitivity to plaintext and ciphertext changes is not achieved. It means that as other stream ciphers, a single bit change on the plaintext only changes one bit on the ciphertext and vice versa. Therefore if the keystream generator was restarted with the same key a successful differential known-plaintext attack could be applied. This fact must be taken into account, above all when continuous IDLE sets are being transmitted over the link.

### E. Security Considerations: Map Reconstruction Attack

One important security problem related with chaotic systems is the possibility of reconstructing the map and inferring the control parameters. This type of attack can be performed by analyzing the output sequence and plotting $x_{i+1}$ versus $x_i$. As an example, if the attacker knows the value of $x_i$ and $x_{i+1}$, he could use (1) for $f(x_{i-1})$ and $f(x_i)$ to calculate the values of $\gamma$ and $x_{i-1}$. Solutions for this kind of attacks are to shuffle/truncate the chaotic orbit before using it for encryption [28].

The fact that we are using only the least 16 significant bits from each $x_i$ for the encryption, limits the information revealed to an attacker. If an attacker could know the least significant 16 bits of $x_i$ and $x_{i+1}$, there would still be $2^{48}$ possible values for each $x_i$ or $x_{i+1}$, then $2^{96}$ for the combination of both.

In addition, the eight least significant bits of each $x_i$ are XORed with the LFSR output which makes the map reconstruction more difficult.

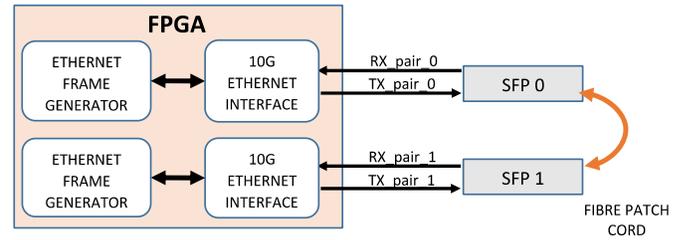

Fig. 10. Test setup scheme.

### F. Security Considerations: Other Considerations

Initial key configuration and key refreshing are important mechanisms in any cryptographic system. In this work, keys have been preconfigured in transmitter and receiver thanks to the FPGA debug system. The implementation of the key management procedure based on 64b/66b control blocks has been let as future work.

As mentioned in Section III, a future set of control messages could be expanded thanks to encryption infrastructure shown in Fig. 3. In order to avoid malicious manipulation or generation of these kinds of messages, a MAC (Message Authentication Control) functionality also should be implemented for the future protocol.

## V. SYSTEM IMPLEMENTATION

### A. System Description

The overall system has been implemented in a Xilinx Virtex 7 FPGA. In the setup for test, the FPGA has been connected to two SFP+ (Small Form-Factor Pluggable) modules capable of transmitting at a rate of 10.3125 Gbps at 850 nm over multimode fiber. The FPGA design consists of two 10G Ethernet interfaces including the 10G-PHYsec function and two Ethernet Frame Generator modules connected to them. The structure of the 10G Ethernet Interfaces is shown in Fig 9. It includes the MAC module and the PHY (PCS and PMA blocks). The PHY is connected directly to the SFP+ modules thanks to the FPGA SERDES (Serializer-Deserializer) circuit. In the MAC side, Ethernet Interface is connected to the Ethernet Frame Generator to test the encrypted link with real traffic. It allows to verify that no frames are lost and no CRC (Cyclic Redundancy Check) errors are produced during encryption.

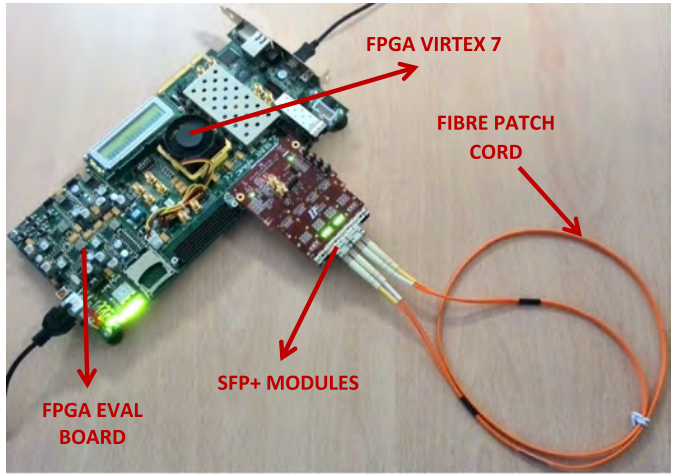

Fig. 11. Test setup photo with SFP+ modules working at 10 Gbps rate.

In Fig. 10 and Fig. 11 the diagram and a photo of the hardware test setup are shown, respectively. Since the key can be configured separately for TX and RX encryption modules in each interface, it is possible to test the bidirectional link with different keys in each direction.

### B. Encryption Results

The conclusions drawn from simulation and hardware debugging can be summarized in the following points:

1) Encryption/Decryption works correctly and synchronously without harming data traffic or link establishment between 10G Ethernet interfaces. Thanks to the frame and CRC counters inside Ethernet Frame Generators it has been possible to check that neither frame losses nor CRC errors are produced. Particularly, bursts of 1024-byte length frames were tested with a duration up to $10^6$ frames and a throughput between 10% and 98% of the maximum line rate.

2) Transmitted frames are indecipherable thanks to encryption. When encryption keys are different between transmitter and receiver, no valid frames are received in the receiver because it is impossible to recover the right 64b/66b block flow.

3) When encrypting, data traffic pattern is indistinguishable from a continuous IDLE pattern, then it is able to hide the pattern of Ethernet traffic from any malicious observer.

For this last capability it is interesting to know the signal behavior at the input of the scrambler and output of the descrambler. Particularly, the block synchronization header can give information about the transmission state.

If the encryption is not enabled, the synchronization header takes the value "10" when transmitting Ethernet frames and "01" during the IFG (Inter Frame Gap) periods and frame boundaries. If no frame is transmitted, the synchronization header is a continuous value "01", as IDLE control characters are being continuously transmitted.

When encryption is enabled, the synchronization header switches randomly between "01" and "10" and the same happens with the 64b/66b block payload that is randomized thanks to the keystream. This happens in both situations, with or without Ethernet traffic being transmitted over the link. This effect makes the data traffic pattern indistinguishable.

TABLE III
FPGA RESOURCES USED BY PCS MODULE

| MODULE | Slice LUTs | Slice Registers | DSPs |
|---|---|---|---|
| PCS (without cipher) | 3601 | 3726 | 0 |
| PCS (with cipher) | 13701 | 8641 | 160 |

TABLE IV
FPGA RESOURCES USED BY ENCRYPTION INFRASTRUCTURE AND KEYSTREAM GENERATORS IN RX AND TX

| MODULE | Slice LUTs | Slice Registers | DSPs |
|---|---|---|---|
| ENCRYP INFRASTR | 276 | 293 | 0 |
| KEYSTREAM RX | 4687 | 2271 | 80 |
| KEYSTREAM TX | 4680 | 2271 | 80 |

TABLE V
FPGA RESOURCES USED IN KEYSTREAM GENERATOR SUBMODULES

| MODULE | Slice LUTs | Slice Registers | DSPs |
|---|---|---|---|
| KEYSTREAM_GEN | 4680 | 2271 | 80 |
| LFSR | 202 | 186 | 0 |
| STM_BANK | 3534 | 1569 | 64 |
| STM_1BIT | 948 | 516 | 16 |

### C. Implementation Results

Regarding the resources used by the PCS layer, it is interesting to study the increase in resources needed to implement the encryption features. In Table III, post-synthesis resource estimation is shown. When encryption is included it is possible to see a large increase in LUTs (Look-Up Table), registers and DSP (Digital Signal Processing) cells. Approximately this increment consists of 10100 LUTs, 4915 registers and 160 DSP cells and is distributed among the RX/TX keystream generators and the encryption infrastructure, as shown in Table IV. In Table V, resources for keystream generator (KEYSTREAM GEN) are decomposed into the 64- bit keystream generator (STM BANK) plus LFSR module used in block payload encryption and the 1-bit keystream generator (STM_1BIT) for block synchronization header encryption.

In the case of DSP cells, 16 of them are necessary to implement the multiplications inside each STM_CELL, as the chaotic map implementation requires it. A total of 80 DSP cells are necessary for the complete keystream generator.

### D. Comparison With Other Solutions

In Table VI a resource and throughput comparison with other chaotic encryption solutions is shown. This comparison

TABLE VI
CHAOTIC CELL COMPARISON WITH OTHER SOLUTIONS

| | Application | Platform | FF | LUTs | DSP cells | Max Freq. (MHz) | Output bits/cycle | Internal state (bits) | Key size (bits) | Encryption Rate (Mbps) | Encryption Rate/Register (Mbps/Register) | Encryption Rate/LUT (Mbps/LUT) | Encryption Rate/DSP (Mbps/DSP) |
|---|---|---|---|---|---|---|---|---|---|---|---|---|---|
| **LM [30]** | PRNG | Virtex-5 | 64 | 129 | 16 | 26.9 | 1 | 64 | 64 | 26.9 | 0.42 | 0.21 | 1.7 |
| **MLM [29]** | Encryption | Virtex-6 | 160 | 643 | 16 | 93 | 16 | 64 | 192 | 1488 | 9.3 | 2.31 | 93 |
| **GFLM [33]** | Encryption | Virtex-5 | 28 | 309 | 3 | 58.4 | 20 | 20 | 60 | 1168* | 41 | 3.77 | 389 |
| **Bernoulli [31]** | PRNG | Spartan 3E | 108 | 575 | 9 | 36.9 | 1/5 | 52 | 156 | 7.38 | 0.07 | 0.01 | 0.82 |
| **Chua [32]** | PRNG | Artix-7 | 787 | 986 | 64 | 80 | 96 | 96 | - | 7685* | 9.76 | 7.79 | 120 |
| **CIPRNG-XOR [34]** | PRNG | Artix-7 | 582 | 345 | 0 | 257.5 | 32 | 32 | - | 8240* | 14.1 | 23.88 | NA |
| **GCIPRNG-F2 [35]** | PRNG | Artix-7 | 444 | 415 | 0 | 205.7 | 32 | 32 | - | 6580* | 14.8 | 15.85 | NA |
| **This Work** | Encryption | Virtex-7 | 64 | 858 | 16 | 175.4 | 16 | 64 | 189 | 2806 | 43.8 | 3.27 | 175 |

*These solutions achieve their Encryption_rate at expense of using their complete state as output, which can reduce the security.

is only illustrative as the FPGA device used in each implementation is different. In this work Virtex-7 is used, however in the other solutions different devices are used, such as Artix-7, Virtex-6, Virtex-5 and Spartan-3E. While CLB structure in Artix-7, Virtex-6 and Virtex-7 is similar in terms of LUTs and registers, with four 6-input LUTs and 8 registers per slice, in Virtex-5 each slice contains four 6-input LUTs and 4 registers. The difference is higher with Spartan-3E, a much older device with only 2 LUTs and 2 register per slice. For this reason the comparison is made in terms of *Encryption_rate/register* and *Encryption_rate/LUT*.

As a comparison among different chaotic cells is a difficult task, we have included in Table VI not only their hardware resources and encryption throughput but also some parameters relative to their structure, such as the internal state and output bit width, and the total key/parameters length. Assuming that every chaotic cell passes the randomness tests, authors consider that the mentioned parameters are important as a measure of the overall security provided by them. Indeed, some of the solutions shown in Table VI are proposed as encryption algorithms while others are used only as PRNGs (Pseudo Random Number Generator) for cryptographic applications, such as key generation.

On the one hand, key size shows how secure can be a cell against brute-force attacks, while the bit width difference between the internal state and output shows how secure can be against a map reconstruction attack. Both parameters affect the results in the hardware resources and encryption rate shown in Table VI. In particular, a longer key and internal state mean that the overall hardware resources are increased, since the width of arithmetic operands inside the cell also increases. In the same way the maximum operation frequency achieved by the cell is decreased, which reduces the final throughput. On the other hand, the opposite effect happens with the output width, a longer output means higher throughput, as more bits per cycle are available for being used in the encryption process.

For example, in [29] the resulting chaotic cell with a 192-bit key obtains its output from the 16 least significant bits of its 64-bit internal state, while in [30], the chaotic cell, also with a 64-bit state, uses only as output one of its bits, however it has a key of only 64 bits. In [31], among the several maps that are implemented, we have selected Bernoulli for the comparsion, it has a 52-bit state and only outputs one bit every 5 clock cycles. Other implementations, as [32], and [33] use their complete state as output, with 96 and 20 bits, respectively. Regarding [34] and [35], they use a chaotic iteration post-processing technique to improve the randomness of linear PRNGs. These do not need DSP cells, however the complete output of the chaotic post-processing is used as output.

According to these parameters, key size, internal state and output bits, the proposed work, together with [29] and [31], presents the best security, with a key of more than 128 bits and less than the 25% of its internal state bits revealed.

In terms of *Encryption_rate* and *Encryption_rate/resource* the implementation in this work clearly achieves a good result. Although [32], [34] and [35] present better *Encryption_rate*, it is at expense of using their complete state as output, which reduces the security. Indeed they are mainly proposed as PRNGs.

Regarding to the DSP cells used, the proposed solution needs a higher quantity than others, as in [31] and [33]. However, it is also at the expense of reducing other parameters. For example in both solutions *Encryption_rate* is lower than this work. Moreover, in [33] the complete state is taken as output and the key size is only 60 bits, which diminishes the security achieved by this solution.

Finally, if we compare this work with [29], the most similar in terms of security, this work achieves better *Encryption_rate* figures. However, it is important to remark that the FPGA used in [29] is a Virtex-6, a 40 nm device, while Virtex-7 used in this work is a 28 nm device. Because of that, the proposed STM cell has been also implemented configuring the target device as the same in [29]. In this case the obtained hardware resources are similar to those used in Virtex-7 but resulting in a lower maximum operation frequency of 154.2 MHz. This fact reduces the *Encryption_rate* but anyway it is still higher that in [29].

## VI. CONCLUSION

As far as the authors are aware, this is the first time that a 10GBase-R Ethernet physical layer encryption method has been proposed and developed. The proposed new encryption function 10G-PHYsec consists of symmetric ciphering at PCS sublayer of the 64b/66b block stream transmitted over an optical link. Encryption based on an original chaotic cipher has been tested with real Ethernet traffic and it has been concluded that the proposed system works correctly without harming data traffic or link establishment, making Ethernet frames indecipherable and obfuscating completely the data traffic patterns. These features improve the security at physical level with no throughput losses, null overhead and low latency, and it is compatible with other solutions at other layers where data packets must be modified by adding extra overhead and delay during their transmission.

Another advantage of the proposed method is that by preserving 64b/66b coding properties as DC-balance, short run length, and transmission density, physical layer encryption is achieved without the necessity of change any of subsequent hardware elements in the TX/RX chain, as SERDES (Serializer/Deserializer), CDR and driver circuits or optical components as commercial SFP modules. In addition, as this method is based in the 64b/66b encoding, it could be scaled to other standards with higher transmission rates based also in this encoding method, such as in 40 or 100 Gigabit Ethernet.

Although it is necessary an increment of FPGA resources for the proposed physical layer modifications, it mainly lies with the keystream generator module and entails an encryption throughput better than other solutions based on chaotic maps.

Finally, we are working on improving the encryption protocol to add functionalities such as new control messages for temporary key refresh and some mechanism for message authentication control.

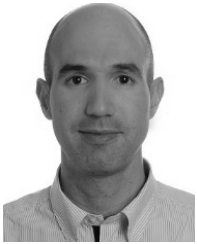

**Adrián Pérez-Resa** was born in San Sebastián, Spain. He received the M.Sc. degree in telecommunications engineering from the University of Zaragoza, Zaragoza, Spain, in 2005.

He is currently pursuing the Ph.D. degree with the Group of Electronic Design, Aragon Institute of Engineering Research, University of Zaragoza.

He was an R&D engineer with telecommunications industry for over 10 years. He is also a member of the Group of Electronic Design, Aragón Institute of Engineering Research, University of Zaragoza. His research interests include high-speed communications and cryptography applications.

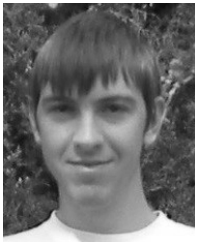

**Miguel Garcia-Bosque** was born in Zaragoza, Spain. He received the B.Sc. and M.Sc. degrees in physics from the University of Zaragoza, Zaragoza, in 2014 and 2015, respectively, where he is currently pursuing the Ph.D. degree with the Group of Electronic Design, Aragon Institute of Engineering Research.

He is also a member of the Group of Electronic Design, Aragón Institute of Engineering Research, University of Zaragoza. His research interests include chaos theory and cryptography algorithms.

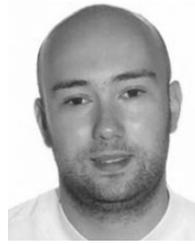

**Carlos Sánchez-Azqueta** was born in Zaragoza, Spain. He received the B.Sc., M.Sc., and Ph.D. degrees in physics from the University of Zaragoza, Zaragoza, in 2006, 2010, and 2012, respectively, and the Dipl.-Ing. degree in electronic engineering from the Complutense University of Madrid, Madrid, Spain, in 2009.

He is currently a member of the Group of Electronic Design, Aragón Institute of Engineering Research, University of Zaragoza. His research interests include mixed-signal integrated circuits, high-frequency analog communications, and cryptography applications.

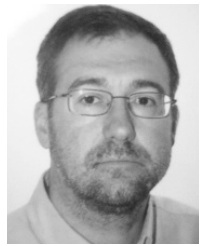

**Santiago Celma** was born in Zaragoza, Spain. He received the B.Sc., M.Sc., and Ph.D. degrees in physics from the University of Zaragoza, Zaragoza, Spain, in 1987, 1989, and 1993, respectively.

He is currently a Full Professor with the Group of Electronic Design, Aragon Institute of Engineering Research, University of Zaragoza. He has co-authored over 100 technical papers and 300 international conference contributions. He has co-authored four technical books and holds four patents. He appears as a principal investigator in over 30 national and international research projects. His research interests include circuit theory, mixed-signal integrated circuits, high-frequency communication circuits, wireless sensor networks, and cryptography for secure communications.